\documentclass[aps,amsmath,amsfonts,twocolumn,showkeys,showpacs]{revtex4}
\usepackage{subfigure}
\usepackage{epsfig}
\usepackage{hyperref}
\newcommand{\ket}[1]{|#1\rangle}
\newcommand{\bra}[1]{\langle #1|}
\newcommand{\braket}[2]{\langle #1| #2 \rangle}

\begin{document}
\title{Spatial non-cyclic geometric phase in neutron interferometry}

\author{Stefan Filipp$^{1,2}$\footnote{Electronic address:
sfilipp@ati.ac.at}, Yuji Hasegawa$^{1}$\footnote{ Electronic
address: hasegawa@ati.ac.at}, Rudolf Loidl$^{1,2}$ and Helmut Rauch$^{1}$}
\affiliation{$^{1}$Atominstitut der \"Osterreichischen
Universit\"aten,
Stadionallee 2, A-1020 Vienna, Austria\\
$^{2}$Institut Laue Langevin, Bo\^ite Postale 156, F-38042 Grenoble
Cedex 9, France} 
\date{\today}
\begin{abstract}
We present a split-beam neutron interferometric experiment to test
the non-cyclic geometric phase tied to the spatial evolution of the
system: the subjacent two-dimensional Hilbert space is spanned 
 by the two possible paths in the
interferometer and the evolution of the state is controlled by phase shifters
and absorbers. A related experiment was reported previously by some of the
authors \cite{hasegawa96} to verify the \emph{cyclic} spatial
geometric phase.
The interpretation of this experiment, namely to ascribe a geometric
phase to this particular state evolution, has met severe criticism \cite{wagh99}. The extension to
\emph{non-cyclic} evolution manifests the
correctness of the interpretation of the previous experiment by means
of an explicit calculation of the 
non-cyclic geometric phase in terms of paths
on the Bloch-sphere. The theoretical treatment comprises
the cyclic geometric phase as a special case, which is confirmed by
experiment.

\end{abstract}
\maketitle

\section{Introduction}
Since the discovery of a geometric effect by Berry
\cite{berry84} in the shape of an additional phase factor after an
adiabatic and cyclic transport of a quantum system Berry's phase has
been intensively investigated and generalized:
the
extension to degenerate subspaces by Wilckzek \cite{wilczek84},
the removal of the adiabatic constraint by Aharonov and Anandan
\cite{aharonov87} and the cyclic condition by Samuel and Bhandari
\cite{samuel88} using the early ideas of Pancharatnam
\cite{pancharatnam56} and the kinematic approach to geomeric phases by
Mukunda and Simon \cite{mukunda93}, to name a few.
In all these contexts the geometric phase is dependent  only on
the geometry of the subjacent Hilbert space, but not on the particular
dynamics of the system under consideration.
Furthermore, Manini and Pistolesi \cite{manini00} proposed an off-diagonal
geometric phase to exhibit the geometry of state space in situations where the usual (diagonal) geometric
phase is undefined. This has been verified experimentally by some of the
authors \cite{hasegawa01}. 

In course of the development of
quantum mechanics it has become clear that the concept of pure states
is not sufficient when taking environmental influences causing
decoherence effects into account. Then one has to use the concept of mixed states. Probably the first
treatise of a geometric phase for mixed states is due to Uhlmann
\cite{uhlmann86} in a quantum algebraic context. Another definition of a
mixed state geometric phase is due to Sj\"{o}qvist \emph{et al.}
\cite{sjoqvist00} using an interferometric approach for its
definition. For both definitions one has to keep in mind that there
exist points in parameter space for which the mixed state geometric
phases remain undefined provoking an extension to off-diagonal mixed
state geometric phases \cite{filipp03a}.

The geometric phase is associated with an evolution of a
system governed by an Hamiltonian, e.~g. a neutron in a
magnetic field where the geometric phase arises by the spinor
evolution due to the coupling with the magnetic field. Here, we observe a geometric phase as an effect of the change
in the spatial degrees of freedom in an interferometry setup. 
A proposal to verify the spatial
geometric phase is due to Sj\"{o}qvist \cite{sjoqvist01} using
polarized neutrons by reversing the r\^oles of the magnetic field and
the spatial degrees of freedom.
Moreover, an experiment using unpolarized neutrons has been performed
by Hasegawa \emph{et al.} \cite{hasegawa96, rauch00} to
test the cyclic spatial geometric phase by inducing a relative
phase shift of $2\pi$ between the interfering neutron beams in a perfect silicon
single-crystal interferometer. The geometric interpretation of this
experiment has been dismissed by Wagh \cite{wagh99} demanding
for further investigations, namely in the non-cyclic case, which is the purpose of the current article.

\section{Geometric phases}
\label{sec:geomphase}

Let us briefly review the basic concepts of geometric phases: A geometric phase is a quantity which is deeply
connected to the curvature of some underlying (state- or parameter-)
space. A two-dimensional plane in three dimensional real space does
not have an intrinsic curvature, but when considering a sphere
embedded in euclidean real space we have to take the curvature of this
manifold into account. In geometry this curvature is reflected for
example in the angle difference of a vector transported around a loop
along geodesics, i.~e. great circles: If a vector is pinned onto a sphere and then
transported along a meridian to the equator, for some angle $\alpha$
along the equator and back to the initial point without changing its
length and its direction in the tangent plane to the surface of the
sphere, the vector will point in a different direction with a
relative angle of $\alpha$ as the \emph{holonomy} associated with the
loop. If we do the same on a two-dimensional plane the initial and the
final vector will point in the same direction.

Berry \cite{berry84} was the first who addressed this issue in quantum
mechanics: He considered a system initially in an eigenstate
$\ket{n(\vec{R}(t))}_{t=0}$ of the governing Hamiltonian $H(\vec{R}(t))$
dependent on the parameters $\vec{R}(t)$ changing with time $t$. As a demonstrative
example one may consider a neutron coupling to a magnetic field
$H(\vec{R}(t))=-\vec{\mu}\cdot\vec{B}(\vec{R}(t))$ due to its
magnetic dipole moment $\vec{\mu} = \mu_n \vec{\sigma}$, where
$\vec{\sigma}=\{\sigma_x,\sigma_y,\sigma_z\}$ are the Pauli
matrices and the magnetic moment $\mu_n$ of a neutron. Suppose now that
the neutron is initially polarized in the direction of the magnetic
field. If the direction of the magnetic field is changed
adiabatically, i.~e. slowly enough to avoid transitions to an
orthogonal state, the system will stay in the eigenstate
$\ket{n(\vec{R}(t))}$ at all times $t$. Furthermore, when tracing out a \emph{loop} in 
parameter space 
the final state $\ket{\Psi(\tau)}$ at time $\tau$ will be the same as the initial state
up to an additional phase factor:
\begin{eqnarray}
  \label{eq:berryphase}
  \ket{\Psi(\tau)}&= &e^{i\phi_d}e^{i\phi_g}\ket{n(\vec{R}(\tau))} \nonumber\\
  &=&e^{-\frac{i}{\hbar}\int_0^\tau E_n((t) dt}e^{-\oint_C
    \bra{n(\vec{R})}\vec{\nabla}_{\vec{R}}\ket{n(\vec{R})}d\vec{R}}\ket{n(\vec{R}(\tau))}.
\end{eqnarray}

The first phase value $\phi_d = {-\frac{1}{\hbar}\int_0^\tau
  E_n(\vec{R}(t)) dt}$ is dependent on the time needed to
transverse the loop and on the energy $E_n(t) =
\bra{\Psi(t)}H(t)\ket{\Psi(t)}$ of the system, whereas the
second phase $\phi_g ={i\oint_C
  \bra{n(\vec{R})}\vec{\nabla}_{\vec{R}}\ket{n(\vec{R})}d\vec{R}}$ is dependent only on the circuit integral in parameter
space revealing the geometric structure. The latter is termed
\emph{Berry phase} or more general \emph{geometric phase}
 in contrast to the former \emph{dynamical phase} $\phi_d$.
$\phi_g$ can be rewritten as a surface integral
 by use of Stoke's Theorem yielding $\phi_g = - \text{Im}\int_F d\vec{S}
 V_n(\vec{R})$, where $F$ is the surface enclosed by the loop in
 parameter space with $d\vec{S}$ denoting the area element and $V_n =
 \nabla \times \braket{n}{\nabla n}$ in an obvious abbreviated
 notation. For the neutron example -- or more
 generally for any
spin-1/2 particle -- $\phi_g$ equals half of the solid angle enclosed by the
loop as seen from the degeneracy point $|\vec{R}| = 0$ in
parameter space. This can also be related to the example from
geometry above where the holonomy after the transport of the vector
pinned initially to the north pole of a sphere
equals the solid angle as seen from the origin of the sphere.

Several restrictions have been relaxed in course of the years,
e.~g. extensions to nonadiabatic \cite{aharonov87}, noncyclic and
nonunitary \cite{samuel88}, and nonpure \cite{sjoqvist00,uhlmann86}
geometric phases have been made. Important in our case are the
generalizations to the nonadiabatic regime by Aharonov
and Anandan and to noncyclic paths by Samuel and Bhandari. 

In this case we have to introduce the Projective Hilbert space (Ray
space) $\mathcal{R}$ by
identifying all state vectors in Hilbert space $\mathcal{H}$ which differ only by an
overall phase factor:
\begin{equation}
\label{equivalence}
\ket{\phi} \sim \ket{\phi'}: \ket{\phi'} = e^{i\alpha} \ket{\phi},\ \alpha \in \mathbb{R}.
\end{equation}
 The stress is therefore
shifted from the parameter space of the Hamiltonian in case of Berry's
construction to state space. We are not interested in the changes of
the driving parameters (as the direction and strength of the magnetic
field) but in the changes of the state itself. In Berry's
considerations these two spaces are identical since the state
follows the changes in parameters due to the adiabaticity condition. 

In the construction by Aharonov and Anandan  one
considers an open path in Hilbert space which is projected to a
path in Ray space by use of the equivalence relation in Eq. (\ref{equivalence}),
i.~e. the curve $\mathcal{C}: t \in [0,\tau] \mapsto \ket{\phi(t)} \in
\mathcal{H}$ is projected to $\widetilde{\mathcal{C}}: t \in [0,\tau] \mapsto
\pi(\ket{\phi(t)})\equiv\ket{\phi(t)}\bra{\phi(t)} \in \mathcal{R}$
with $\ket{\phi(0)}\sim\ket{\phi(\tau)}$. For $\widetilde{\mathcal{C}}$ an absolute phase factor of $\ket{\phi(t)}$ is
immaterial, since the curve $\widetilde{\mathcal{C}}$ is defined via the
evolution of the projection operator $\ket{\phi(t)}\bra{\phi(t)}$. The geometric phase
is a property of Ray space, where $\widetilde{\mathcal{C}}$ is closed
due to the equivalence of the initial and the final state
$\ket{\phi(0)}\sim\ket{\phi(\tau)}$ and can be calculated via a surface integral over the area
enclosed by  $\widetilde{\mathcal{C}}$. 

One can find many different curves
$\mathcal{C}',\mathcal{C}'',\ldots$ in
Hilbert space differing by a phase factor $e^{i\alpha(t)}$ and
yielding the
same curve in Ray space under the projection map $\pi$. On the other hand for a given curve in Ray
space there exists one distinct curve in Hilbert space fulfilling the parallel
transport conditions, namely that two neighbouring states
$\ket{\phi(t)}$ and $\ket{\phi(t+dt)}$ in $\mathcal{H}$ have the same phase, that is to
say, $\braket{\phi(t)}{\phi(t+dt)}$ is real and positive. This
implies by Taylor expansion that
$\bra{\phi(t)}\frac{d}{dt}\ket{\phi(t)}=0$ \footnote{These are natural
  parallel transport conditions analogue to the claim for constant
  lenght and direction of a parallel transported vector on a sphere in
  the example from geometry above.}. For this curve the dynamical
phase vanishes as one can verify by inserting the Schr\"odinger
equation $H(t)\ket{\phi(t)}=i\hbar d/dt\ket{\phi(t)}$ into the parallel transport condition.

The concept can be extended to apply to open paths in Ray space where $\ket{\phi(\tau)}\not\sim \ket{\phi(0)}$ by closing the curve by
a geodesic, i.~e. a path in Ray space with the shortest distance from $\ket{\phi(\tau)}\bra{\phi(\tau)}$ to
$\ket{\phi(0)}\bra{\phi(0)}$. Then one obtains a well-defined
surface area enclosed by the path generated by the evolution of the
system plus the geodesic closure. This surface provides an expression
for the geometric phase, which has been proven by Samuel and
Bhandari \cite{samuel88}.

To sum up, for a general
evolution of a quantum state the state obtains a dynamical phase
dependent on the energy and time as well as a geometric phase only dependent on
the subjacent geometry of state space. For special Hamiltonians which
fulfill the parallel transport conditions the dynamical phase
vanishes, which is also the case when the state is transported
along a geodesic. An example of the latter is an evolution along a great
circle on a sphere for a two-level system which we will encounter in
the forthcoming discussion.

\section{Interferometric setup}
Due to Feynmann \cite{feynman57} the description of any two-level
quantum system is equivalent to the description of a spin-1/2
particle. Exploiting this equivalence there is in principle no
difference between manipulations in the spin space of neutrons with
the orthogonal basis $\{\ket{\uparrow},\ket{\downarrow}\}$ as
eigenstates of $\sigma_z$, and momentum
space with $\{\ket{k},\ket{k'}\}$ as orthogonal basis vectors
corresponding to two directions of the neutron beam in an
interferometer. In both cases one can assign a geometric phase
to the particular evolution of the initial state. An even more appropriate description for the
interferometric case for the
forthcoming discussion is in terms of ``which-way'' basis states
$\{\ket{p},\ket{p^\perp}\}$, namely, if the neutron is found in the upper
beam path after a beamsplitting plate it is said to be in the state $\ket{p}$,
or in the state $\ket{p^\perp}$, if found in the lower beam path. In case
of a $50:50$ beamsplitting of the incident
(neutron) beam into a transmitted beam and a reflected beam
 the associated
wave vector after the beamplitter can be written as 
an equally weighted coherent superposition of the two paths
$\ket{q(\delta)}\equiv1/\sqrt{2}(\ket{p}+ e^{i \delta} \ket{p^\perp})$ with the relative
phase $\delta \in \mathbb{R}$ depending on the particular physical
realization of the beamsplitter.

\begin{figure}[htbp]
  \centering
  \includegraphics[width=80mm]{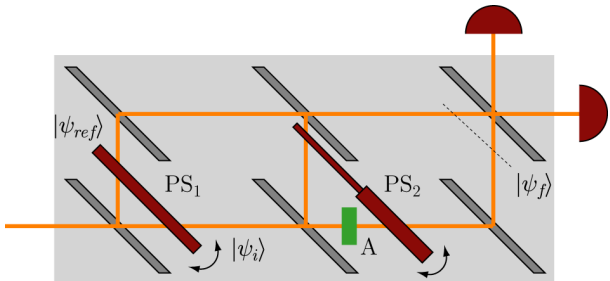}
  \caption{Experimental setup to test the spatial geometric phase in a neutron interferometer}
  \label{fig:2004-splitbeamNIST-setup}
\end{figure}
For testing the spatial geometric phase we use a double-loop
interferometer (cf. Fig. \ref{fig:2004-splitbeamNIST-setup}), where the incident unpolarized
neutron beam $\ket{\psi}$ is split up into a diffracted reference beam $\ket{\psi_{ref}}$ and
a transmitted beam $\ket{\psi_t}$. 
The
transmitted beam is subjected to further
evolution in the second loop of the interferometer by use of
beamsplitters (BS1 and BS2), an absorber (A)
with transmission coefficient $T$ and the  phase shifter
PS2 generating a phase shift of $e^{i\phi_1}$ on the upper and
$e^{i\phi_2}$ on the lower beam path,
respectively, yielding the final state $\ket{\psi_f} =
U\ket{\psi_t} = U \ket{p}$. Here $\ket{\psi_t}=\ket{p}$ since before
the beam splitter BS1 the beam is clearly localized as seen from
the second loop of the interferometer. The unitary matrix $U=U(T,\phi_1,\phi_2)$ comprises all the manipulations
in the second loop:
\begin{eqnarray}
  \label{eq:evolution}
  \ket{\psi_t} & \xrightarrow{\text{BS}}& 
  \frac{1}{\sqrt{2}}(\ket{p}+\ket{p^\perp}) \xrightarrow{\text{A}}
    \frac{1}{\sqrt{2}}(\ket{p}+\sqrt{T}\ket{p^\perp})\nonumber\\
    &\xrightarrow{\text{PS2}}&\frac{1}{\sqrt{2}}(e^{i\phi_1}\ket{p}+\sqrt{T}e^{i\phi_2}\ket{p^\perp}) \equiv U\ket{\psi_t} = \ket{\psi_f}.
\end{eqnarray}
The geometric phase can then be extracted from the argument of the
complex valued
scalar product between the initial and the final state $\arg
\braket{\psi_t}{\psi_f}$ (when removing dynamical contributions as will
be discussed later). This is where the reference beam comes into play:
$\ket{\psi_{ref}}$ is not subjected to any further evolution, but is
stationary apart from adding a phase factor $e^{i\eta}$ by use of the
phaseshifter PS1. $\ket{\psi_{ref}}$ propagates towards the beamsplitter
BS2 from the upper path, thus we can assert it to be in the state
$e^{i\eta}\ket{p}$. Then by the variable
phase shift $e^{i\eta}$  one can measure the
shift of the interference fringes reflecting the phase difference
between $\ket{\psi_{ref}}$ and $\ket{\psi_f}$.


This preparation of the states is followed by the recombination of the two beams $\ket{\psi_f}$ and
$\ket{\psi_{ref}}$ at the beamsplitter BS2
 and the detection at the detector D$_O$ in the forward
beam. This step can be described by the application of the projection
operator $\ket{q}\bra{q}=1/2(\ket{p}+\ket{p^\perp})(\bra{p}+\bra{p'})$ 
(with $\delta=0$, which can always be achieved by an appropriate choice
of the phase of the basis states) to $\ket{\psi_f}$ as well as to
$\ket{\psi_{ref}}$:
  \begin{eqnarray}
    \ket{\psi_f^\prime}&=&\ket{q}\braket{q}{\psi_f}=K(e^{i\phi_1}+\sqrt{T}e^{i\phi_2})\ket{q}\nonumber\\
    \ket{\psi_{ref}^\prime}&=&\ket{q}\braket{q}{\psi_{ref}} =K\ket{q},
  \end{eqnarray}
where $K$ is some scaling constant. 

The intensity $I$ measured in the detector $D_O$ is proportional to the absolute square of the
superposition $\ket{\psi_f^\prime}+ e^{i\eta}\ket{\psi_{ref}^\prime}$:
\begin{eqnarray}
  \label{eq:interference}
  I&\propto&\big|(e^{i\eta} + e^{i\phi_1} \sqrt{T}e^{i\phi_2}) \ket{q}\big|^2 = \langle \psi_{ref}^\prime
  | \psi_{ref}^\prime \rangle + \langle \psi_f^\prime | \psi_f^\prime \rangle +\nonumber\\
&& + 
2 |\langle \psi_{ref}^\prime | \psi_f^\prime \rangle |\cos\left( \eta - \arg \langle
\psi_{ref}^\prime | \psi_f^\prime \rangle\right).
\end{eqnarray}
We notice a phase shift of the interference pattern by $\arg
\langle \psi_{ref}^\prime | \psi_f^\prime \rangle$. This phase shifts corresponds to
the Pancharatnam connection \cite{pancharatnam56} between the state
$\ket{\psi_f^\prime}$ and the state $\ket{\psi_{ref}^\prime} = \ket{q}\braket{q}{\psi_t}=\ket{q}$ from
which we can extract the geometric phase. 
Explicitly we obtain 
\begin{eqnarray}
  \label{eq:phase}
  \phi &=& \arg  \langle \psi_{ref}^\prime | \psi_f^\prime \rangle\nonumber\\
  &=& \frac{\phi_1+ \phi_2}{2} -
  \arctan\left[\tan\left(\frac{\Delta\phi}{2}\right)\left(\frac{1-\sqrt{T}}{1+\sqrt{T}}\right)\right],
\end{eqnarray}
where $\Delta\phi\equiv\phi_2 - \phi_1$.
The geometric phase is defined as \cite{mukunda93}
\begin{equation}
  \label{eq:geomdef}
  \phi_g \equiv \arg  \langle \psi_{ref}^\prime | \psi_f^\prime \rangle - \phi_d,
\end{equation}
where $\phi_d$ denotes the dynamical part. 
From Refs. \cite{hasegawa96} and \cite{wagh99} we know that the dynamical part
stemming from the phase shifter PS2 is given by a weighted sum of
the phase shifts $\phi_1$ and $\phi_2$ with the weights depending on
the transmission coefficient $T$. In particular we have
\begin{equation}
  \label{eq:dynphase}
  \phi_d = \frac{\phi_1 + T\phi_2}{1+T},
  \end{equation}
which vanishes by an appropriate choice of phase shifts and
transmission, i.~e. $\phi_d = 0$ for $\phi_1/\phi_2 = -T$.

\begin{figure}[htbp]
  \centering
  \includegraphics[width=70mm]{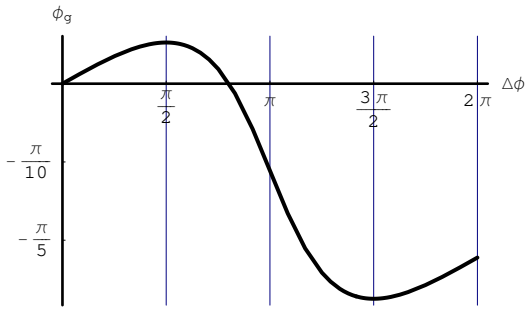}
  \caption{Geometric phase $\phi_g\,\mbox{[rad]}$ in dependence on the
    relative phase shift $\Delta\phi\,\mbox{[rad]}$ for $T=1/8$.}
  \label{fig:2004-splitbeamNIST-phaseshift}
\end{figure}

By varying the relative phase $\Delta\phi$ from $0$ to $2\pi$ and
setting $\phi_d=0$ the geometric phase $\phi_g$ can be plotted over
$\Delta\phi$ (cf. Figure \ref{fig:2004-splitbeamNIST-phaseshift}).

\section{Bloch-sphere description}
For every two-level system we can use the \emph{Bloch-sphere} for 
depicting the state vectors
and evolutions thereof as points and curves on a sphere. Then the
results obtained above, i.~e. the shift of the interference pattern in
Eq. (\ref{eq:interference}) without dynamical contributions, should be
equal to the (oriented) surface area enclosed by the paths of the state vectors
on the \emph{Bloch-sphere}, or, equivalently, to the solid angle
traced out by the state vectors as seem from the origin of the sphere.

\begin{figure}[htbp]
  \centering
  \includegraphics[width=60mm]{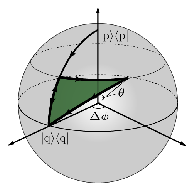}
  \caption{Path of the state in an interferometer on the Bloch sphere
    representing the 2-level system (upper path $\ket{p}\bra{p}$ and
    lower path $\ket{p^\prime}\bra{p^\prime}$).}
  \label{fig:2004-splitbeamNIST-bloch}
\end{figure}

As we can observer
in Figure \ref{fig:2004-splitbeamNIST-bloch} the north
pole of the sphere can be identified with a state with well known path, i.~e. an
eigenstate of the observable $\ket{p}\bra{p}$. After the beam splitter
BS1 the state $\ket{\psi_t}$ evolves to an equal superposition of upper
path and lower path, therefore the evolution on the bloch sphere is
given by a geodesic from the north pole to the equatorial line (the
particular point on the equator is arbitrary due to the arbitrary
choice of the phases of the basis vectors). 

The absorber changes the weights of the superposed basis states, in
particular for the extremal values of $T$ parameterized by the angle
$\theta$ with $T=\tan^2\theta/2$ we end up either again with
an equally weighted superposition for no absorption ($T=1$ or $\theta=\pi/2$) or the
state is now on the north pole for total absorption ($T=0$ or $\theta=0$), since in
the latter scenario we know the particle has taken the upper path when
detecting a neutron in $D_O$. For $T\in (0,1)$ the state is encoded as
a point on the geodesic from the north pole to the equatorial line.

Due to the phase shifter PS2 we obtain a relative phase shift
between the superposing states of $\Delta\phi = \phi_2-\phi_1$:
\begin{equation}
  \frac{1}{\sqrt{2}}(\ket{p}+\ket{p^\perp}) \mapsto \frac{1}{\sqrt{2}}(\ket{p}+e^{i\Delta\phi}\ket{p^\perp}).
\end{equation}
This can be depicted as an evolution along a circle of latitude on the Bloch
sphere with perodicity of $2\pi$.

The recombination at BS2 followed by the detection of the forward beam
in D$_O$ is represented as a projection to the
starting point on the equatorial line, i.~e. we have to close the
curve associated with the evolution of the state by a geodesic 
to the point $\ket{q}\bra{q}$ on the sphere as discussed for
non-cyclic paths in Section \ref{sec:geomphase}.
As for the reference state $\ket{\psi_{ref}}$ we note that the phase
shift of $\eta$ has no impact on the position of the state on the
Bloch sphere, it stays at the north pole. Due to the recombination at
BS2 and the detection the
state is also projected to $\ket{q}\bra{q}$ contributing to the
forward beam incident to the detector D$_O$.

\begin{figure}[htbp]
  \centering
\subfigure[{ Cyclic evolution}]{\label{fig:2004-splitbeamNIST-bloch360}\includegraphics[width=60mm]{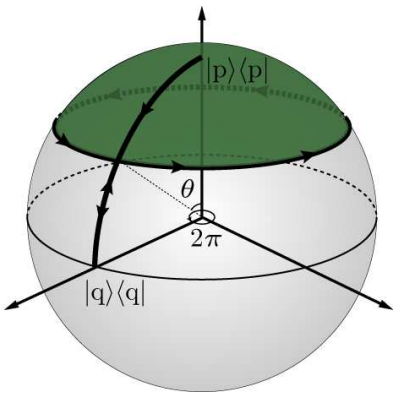}}
\subfigure[{ Non-cyclic evolution}]{\label{fig:2004-splitbeamNIST-bloch180}\includegraphics[width=60mm]{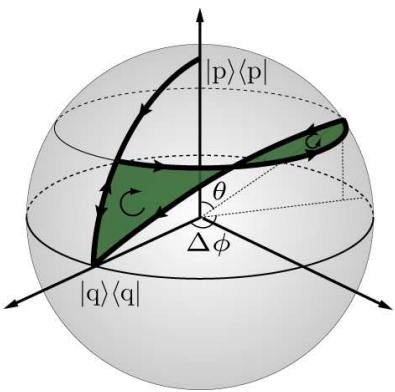}}
  \caption{paths on the Blochsphere corresponding to the evolution of the state in the splitbeam experiment}
  \label{fig:2004-splitbeamNIST-paths}
\end{figure}

The paths are depicted in Figure \ref{fig:2004-splitbeamNIST-paths} in
detail for cyclic \ref{fig:2004-splitbeamNIST-bloch360} as well as non-cyclic evolution \ref{fig:2004-splitbeamNIST-bloch180}. For a relative phase difference greater than $\pi/2$ we have to
take the direction of the loops into account. In \ref{fig:2004-splitbeamNIST-bloch180} the first loop is
transversed clockwise, whereas the second loop is transversed
counter-clockwise yielding a positive or negative contribution to the
geometric phase, respectively.

The numerical calculation of the surface area $F$ enclosed by the path
transversed by the neutron is straightforward by evaluating the
solid angle  $\Omega =\int_F \sin\theta\,d(\!\Delta\phi) d\theta$ via
a surface integral and using $\phi_g = -\Omega/2$. For the cyclic case this integral can be solved easily 
 by calculating the segment on the
sphere according to Figure \ref{fig:2004-splitbeamNIST-bloch360} to obtain
$\phi_g = -\Omega/2=\pi(cos\theta-1)$. For the non-cyclic case no analytic expression has been found to compare the results with the
phase shift of the interference fringes appearing in
Eq. (\ref{eq:interference}). However, the numerical results are
equivalent and agree with Figure \ref{fig:2004-splitbeamNIST-phaseshift}.

\section{Experimental results}

\begin{figure}[htbp]
  \centering
  \includegraphics[width=80mm]{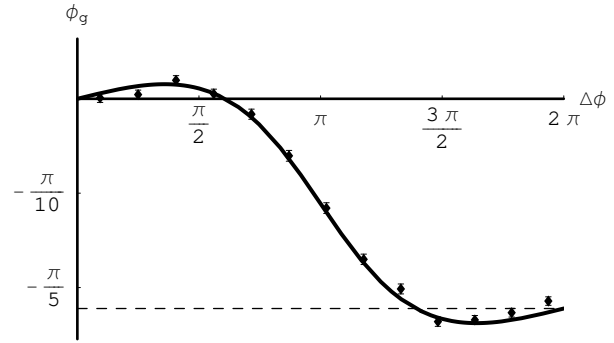}
  \caption{Experimental verification of the spatial geometric phase using a neutron interferometry setup}
  \label{fig:2004-splitbeamNIST-phaseshiftexp}
\end{figure}

For an experimental test of the spatial geometric beam we have used
the double-loop perfect-crystal-interferometer installed at the S18-beamline
at the high-flux reactor ILL, Grenoble \cite{zawisky02}. A schematic
view of the setup is shown in Figure \ref{fig:2004-splitbeamNIST-setup}. Before
falling onto the skew-symmetric interferometer the incident neutron
beam was collimated and monochromatized by the 220-Bragg reflection of
a Si perfect crytal monochromator placed in the thermal neutron guide
H25. The wavelength was tuned to give a mean value of $\lambda_0 =
2.715 \text{\AA}$. To eliminate the higher harmonics we have used
prism-shaped silicon wedges. The beam cross-section was confined to
$5\times 5 \text{mm}^2$ and an isothermal box enclosed the
interferometer to achieve reasonable thermal environmental isolation.
For the phase shifters parallel sided aluminium plates have been used:
$4$mm inserted in the first loop as PS1 and $4$mm and $0.5$mm,
respectively, in the second loop for PS2 yielding a ratio of
$1/8$ for $\phi_1/\phi_2$. To avoid dynamical phase contributions a gadoldinium solution
with $T=0.118$ \footnote{The deviation from the ratio $1/8$ is due to
  the different absorptions of the $4$mm and $0.5$mm Al-phaseshifters.} has been used as an absorber.
For a comparison with the theoretically predicted values one has to
keep in mind that the contrast reflecting the coherence properties is
different between each of the beams in the
interferometer. Accounting for
this experimental fact in the theoretical derivation of the geometric
phase we notice a slighlty flattened curve in Figure
\ref{fig:2004-splitbeamNIST-phaseshiftexp} compared to Figure
\ref{fig:2004-splitbeamNIST-phaseshift}. Nevertheless, one can
recognize
 the increase in geometric phase for $\Delta\phi \in
[0,\pi/2]$ due to the positively oriented surface followed by a decrease
due to appearance of a counter-clockwise transversed loop on the
sphere yielding a negative phase contribution. 

\section{Conclusions}
In summary we have shown that one can ascribe a geometric phase not
only to spin evolutions of neutrons, but also to evolutions in the
spatial degrees of freedom of neutrons in an interferometric
setup. This equivalence is evident from the description of both cases via
state vectors in a two dimensional Hilbert space. However, there have
been arguments contra the experimental verification in
\cite{hasegawa96} which we believe
can be settled in favour of a geometric phase appearing in the setup
described above. The twofold calculations of the geometric either in
terms of a shift in the interference fringes or via surface integrals
in an abstract state space allows for a geometric interpretation of
the obtained phase shift. 

\section*{Acknowledgments}
The authors acknowledge support from the Austrian Science Foundation,
Projects No. F1513 and F1514.

\end{document}